\documentclass[11pt]{article}
\usepackage{latexsym}
\def\appendix{\renewcommand{\thesection}{\Alph{section}}\setcounter{section}{0}
              \renewcommand{\theequation}
            {\mbox{\Alph{section}.\arabic{equation}}}\setcounter{equation}{0}}

\def\maketitle{\thispagestyle{empty}\setcounter{page}0\newpage
                \renewcommand{\thefootnote}{\arabic{footnote}}
                  \setcounter{footnote}0}
\renewcommand{\thanks}[1]{\renewcommand{\thefootnote}{\fnsymbol{footnote}}
               \footnote{#1}\renewcommand{\thefootnote}{\arabic{footnote}}}

\renewcommand{\title}[1]{\begin{center}\Large\bf #1\end{center}\rm\par\bigskip}
\renewcommand{\author}[1]{\begin{center}\Large #1\end{center}}

\newcommand{\pacs}[1]{\smallskip\noindent{\sl PACS numbers:
                       \hspace{0.3cm}#1}\par\bigskip\rm}
\def\babs{\hrule\par\begin{description}\item{Abstract: }\it} 
\def\eabs{\par\end{description}\hrule\par\medskip\rm}
\renewcommand{\date}[1]{\par\bigskip\par\sl\hfill #1\par\medskip\par\rm}
 
\def\beq{\begin{eqnarray}}    
\def\eeq{\end{eqnarray}}      
\def\at{\left(}               
\def\ct{\right)}              
\def\R{{\hbox{{\rm I}\kern-.2em\hbox{\rm R}}}}   
\def\H{{\hbox{{\rm I}\kern-.2em\hbox{\rm H}}}}   
\def\N{{\hbox{{\rm I}\kern-.2em\hbox{\rm N}}}}   
\def\C{{\ \hbox{{\rm I}\kern-.6em\hbox{\bf C}}}} 
\def\Z{{\hbox{{\rm Z}\kern-.4em\hbox{\rm Z}}}}   

\begin{document}

\title{On the \( \sqrt{2} \) puzzle in AdS\(_{2}\)/CFT\(_{1}\)}
\author{G. Catelani\(^{a,b}\)\thanks{catelani@grad.physics.sunysb.edu} , 
L. Vanzo\(^{a}\)\thanks{vanzo@science.unitn.it} \\
\small{\(^{a}\)Dipartimento di Fisica, Universit\`a di Trento\\
and Istituto Nazionale di Fisica Nucleare,\\
Gruppo collegato di Trento, Italia}\\ 
\small{\(^{b}\)Department of Physics and Astronomy,\\ State University of New York at
Stony Brook}}


\begin{abstract}
In this letter we analyze the Hamiltonian formulation of the
Jackiw-Teitelboim model of 2D gravity and calculate the central charge
associated with the asymptotic symmetries, taking care of boundary
terms. For black hole solutions, we show that there is no \(\sqrt{2}\)
discrepancy between the thermodynamical entropy and the statistical
one obtained via Cardy's formula. 
\end{abstract}

\pacs{04.20.-q, 04.70.-s, 04.70.Dy}
\bigskip

\noindent Two-dimensional dilatonic models of gravity received
recently much attention. Their importance relays on their role in the
conjectured AdS/CFT correspondence \cite{cm1,cm2} and in describing some
properties of higher-dimensional black-holes (see
e.g. \cite{nns}). Two distinct realizations of the AdS\(_{2}\)/CFT
correspondence are known \cite{ccav}. The simplest one is known as
AdS\(_{2}\)/CFT\(_{1}\) and can be understood in terms of boundary
fields \cite{nns}, but is believed to be plagued by a value of the
central charge that leads to a mismatch between statistical and
thermodynamical entropy of black hole solutions of the
Jackiw-Teitelboim model. In a previous paper \cite{ccv} we have shown
that if the $U(1)$ timelike symmetry of a stationary solutions has 
fixed points, then the hamiltonian for gravity has a boundary term
coming from the fixed point set, and moreover that this term can be
interpreted in a CFT language. This gave a statistical entropy
matching correctly to the usual thermodynamics of black holes. It was
clear from \cite{ccv} that this horizon boundary term is a general
feature of gravity in any dimension.\\
Take for example the BTZ\cite{btz} black hole, with metric
\begin{equation}
ds^2=-\at-8mG_3+\frac{r^2}{\ell^2}\ct 
dt^2+\at-8mG_3+\frac{r^2}{\ell^2}\ct^{-1}dr^2+r^2d\phi^2
\end{equation}
We normalize the surface gravity to one, so $\sqrt{8mG_3}=\ell$ and the
euclidean metric describes a disk times a circle, both with periodicity
$2\pi$. The mass, i.e. the boundary term at infinity, is then
$m=\ell^2/8G_3$ and the boundary term at the horizon is evaluated as
$\ell^2/4G_3$. As explained in \cite{ccv}, this would give a central
charge $c=3\ell^2/G_3$, $L_0=m$ and the entropy takes the value
\begin{equation}
S=2\pi\sqrt{\frac{cL_0}{6}}=\frac{\pi\ell^2}{2G_3}=\frac{A}{4}
\end{equation}
matching the Bekenstein-Hawking's entropy. On the other hand,
Strominger\cite{stro98} had already found the correct entropy by using
the central charge associated with the asymptotic symmetries in
$AdS_3$ at infinity, as given by Brown-Henneaux\cite{bh86} years
ago. Note however that Brown-Henneaux's central charge is different
from our $c=3r_+/G_3$, which depends from the mass of the black
hole. This would mean that the central charge actually depends on the
states on which the Virasoro operators act. The coincidence of the
present calculation with that of Strominger suggest that the boundary
CFT has enough information to describe the black holes in the bulk
\cite{bbo99}, certainly an aspect of the $AdS_3/CFT_2$ correspondence.\\ 
The aim of this paper is to show that the same happens to the central
charge associated with the asymptotic symmetries of AdS\(_{2}\), that has
indeed the value needed to account for black hole entropy.\\
As a starting point let's consider the JT action\footnote[1]{see
\cite{cm1,cm2,nns} for references, definitions and notation. We try to
summarize the points needed to make our discussion self-contained}
\begin{equation}
\label{action}
S=\frac12 \int d^{2}x \sqrt{-g}\eta[R+2\lambda^{2}]
\end{equation}
The general stationary solution to the equations of motion are given by
\begin{equation}
\label{metric}
ds^{2}=-(\lambda^{2}x^{2}-a^{2})dt^{2}+(\lambda^{2}x^{2}-a^{2})^{-1}dx^{2}
\end{equation}
for the metric and
\begin{equation}
\label{dilaton}
\eta=\eta _{0}\lambda x
\end{equation}
for the dilaton. For positive \(a^{2}\), these solutions can be interpreted 
as black holes of mass
\begin{equation}
\label{mass}
M=\frac12 \eta _{0} \lambda a^{2}
\end{equation}
and entropy
\begin{equation}
\label{tdentr}
S=2 \pi \eta _{0}a
\end{equation}
The metric is defined to be asymptotically AdS\(_{2}\) if, for 
\(x \rightarrow \infty \)\begin{eqnarray}
g_{tt} &=& -\lambda^{2}x^{2}+\gamma_{tt}(t)+O \left( \frac{1}{x^{2}} \right) \nonumber \\
g_{tx} &=& \frac{\gamma_{tx}(t)}{\lambda^{3}x^{3}}+O\left(\frac{1}{x^{5}}\right) \nonumber \\
g_{xx} &=&
\frac{1}{\lambda^{2}x^{2}}+\frac{\gamma_{xx}(t)}{\lambda^{4}x^{4}}+O\left(
\frac{1}{x^{6}} \right)  \nonumber
\end{eqnarray}
while the asymptotic behaviour of the dilaton is taken to be
\[
\eta = \eta_{0} \left(\lambda \rho(t) x+ \frac{\gamma_{\phi
\phi}}{2\lambda x} \right)+O\left( \frac{1}{x^{3}} \right)
\]
The boundary fields are then \( \gamma_{tt}, \gamma_{tx}, \gamma_{xx},
\gamma_{\phi \phi}, \rho \) and they transform as conformal fields
under the action of the asymptotic symmetries. The task is to
calculate the charge J associated with these symmetries and this can
be easily done in the Hamiltonian formalism.\\
Hamiltonian formulation of 2D dilatonic gravity theories has been
already carried out, e.g. in \cite{lgk,krv97}, and other authors
usually refer to that analysis. The starting point is the ADM
decomposition of the metric
\[
ds^{2}=-N^{2}dt^{2}+\sigma^{2}(dx+N^{x}dt)^{2}
\]
which leads to an Hamiltonian of the following form
\begin{equation}
\label{hb}
H_{bulk}=\int dx (N \mathcal{H}+N^{x} \mathcal{H}_{x})
\end{equation}
As explicitly stated in \cite{lgk}, in this expression some boundary
terms have been dropped. We already noticed in \cite{ccv} that the
full Hamiltonian must include this boundary terms and hence it reads
\begin{equation}
\label{ht}
H_{tot}=H_{bulk}+\lim_{x \rightarrow \infty} \left[ \left( \frac{\eta N'}{\sigma}-\eta\Pi_{\eta}N^{x} \right) + \left( \sigma\Pi_{\sigma}N^{x}-\frac{\eta'N}{\sigma} \right) \right]
\end{equation}
where a prime denotes differentiation w.r.t. x and the two \(\Pi\)'s
are the canonically conjugate momenta to the fields in the subscript.\\
The two boundary terms have different origins. The first one comes
from an integration by parts performed in the action integral to get
rid of second order derivatives; the second one arises as a
consequence of a second integration by parts performed after the
Legendre transformation in order to write the Hamiltonian in the form
shown above, with \(N\) and \(N^{x}\) acting as Lagrange multipliers.\\
It is usually believed that there's no need to keep track of these
boundary terms, since we should be able to recover them by requiring
the final Hamiltonian to be differentiable. Using this approach, the
boundary term that has to be added to the variation of (\ref{hb}) is
\[
\delta J=-\lim_{x \rightarrow \infty} [N(\sigma^{-1}\delta\eta'-\sigma^{-2}\eta'\delta\sigma)-N'(\sigma^{-1}\delta\eta)+N^{x}(\Pi_{\eta}\delta\eta-\sigma\delta\Pi_{\sigma})]
\]
or using the boundary fields
\begin{equation}
\label{djb}
\delta J [ \epsilon ] = \eta_{0} \left[ \lambda \epsilon ( \gamma_{tt} \delta \rho + \frac{\rho}{2} \delta \gamma_{xx}+\delta \gamma_{\phi \phi} ) + \frac{\dot{\epsilon} \delta \dot{\rho}}{\lambda} - \frac{\ddot{\epsilon} \delta \rho}{\lambda} \right]
\end{equation}
where \(\epsilon(t)\) is the function of time which enters into the definition of the asymptotic symmetries and a dot denotes differentiation w.r.t. time.
Moreover in \cite{cm2} it's claimed that the equation of motion of the
dilaton (the 00 and 01 component respectively) gives the constraints
\begin{eqnarray}
\label{cons1}
\lambda^{-2}\ddot{\rho}=\rho (\gamma_{tt}-\gamma_{xx} )- \gamma_{\phi \phi} \\
\label{cons2}
\dot{\rho} \gamma_{tt} + \frac{\rho}{2}\dot{\gamma}_{xx} + \dot{\gamma}_{\phi \phi} = 0
\end{eqnarray}
Using these equations and expanding near the classical solution
\(\rho=1+\overline{\rho} \), their value for the charge for on-shell
field configurations, up to a constant which equals the mass
(\ref{mass}) and up to a total time derivative (which plays no role,
since the actual charge is obtained by integrating the above
expression over time \cite{cm1}), is
\begin{equation}
\label{jbf}
J[ \epsilon ] = -2\eta_{0}/\lambda\;\epsilon\ddot{\rho} + \epsilon M
\end{equation}
In the spirit of AdS/CFT correspondence, this charge (once we drop the
mass term) is interpreted as a conformal stress-tensor, whose anomaly
in the conformal transformation gives the central charge of the
associated Virasoro algebra. The value of the central charge obtained
from (\ref{jbf}) is claimed to be \( C=24\eta_{0} \). 
This calculation has two shortcomings. First, in obtaining (\ref{djb})
the boundary terms have been discarded. Second, the use of
(\ref{cons2}) is controversial: equation (\ref{cons1}) must
definitively hold, since it's obtained by requiring the vanishing of
the \(O(x)\) term in the $00$ component of the dilaton equation of
motion; equation (\ref{cons2}), on the other hand, comes from the
leading term of \(O(1/x^{2}) \) in the $01$ component, which therefore
becomes a null identity as \(x\) goes to infinity. Third, to obtain a
Virasoro's algebra the mass term has to be dropped, but setting
\(M=0\) (i.e. \(L_0=0\)) Cardy's formula cannot be used, since it
holds if \(L_0\gg C\).\\
The boundary term in (the phase space functional) \(H_{tot}\)
(\ref{ht}) can be rewritten as
\[
\eta_{0} [ \lambda \epsilon ( \gamma_{tt} \rho+ \gamma_{\phi \phi} ) +\dot{\epsilon}\dot{\rho}/\lambda - \rho \ddot{\epsilon}/\lambda  ]
\]
Including this term in the variation leads to a modified expression for \(\delta J\)
\[
\delta J_{tot} [ \epsilon ] = \eta_{0}\lambda\epsilon\rho [ \delta \gamma_{tt}-\delta\gamma_{xx}/2 ]
\]
This can be rewritten in the more useful form
\begin{equation}
\delta J_{tot} [ \epsilon ] = \eta_{0}\lambda\epsilon \{ \delta [\rho ( \gamma_{tt}-\gamma_{xx} ) ] + \delta\gamma_{xx}/2-\delta\rho ( \gamma_{tt}-\gamma_{xx} ) \}
\end{equation}
In this form, we can immediately insert (\ref{cons1}).
Considering as above configurations near the classical one it's  easy to get
\begin{equation}
J_{tot} = \eta_{0}/\lambda\; \epsilon \ddot{\overline{\rho}} + \epsilon M
\end{equation}
This gives a conformal stress-tensor exactly 1/2 of the previous
result (\ref{jbf}), modulo a sign, and we expect the central charge
also to get a 1/2 factor: \( \mathcal{C} = 12\eta_{0} \).\\
To verify this, we calculate the variation of the stress-tensor
(keeping the mass term!); using $(14)$, $(11)$ and the transformation
laws for the fields given in \cite{cm2}, we get near the classical solutions:
\[
\epsilon\delta_{\omega}\Theta=\epsilon \left( \omega \dot{\Theta}+2\dot{\omega}\Theta \right)-\frac{\eta_0}{\lambda}\,\epsilon\stackrel{\cdots}{\omega}\]
and this equals the Dirac bracket between the charges. Using the Fourier expansions:
\[ 
\epsilon= \frac{1}{\lambda} \sum_{m} a_{m}e^{i\lambda mt},\; \omega= \frac{1}{\lambda} \sum_{n} a_{n}e^{i\lambda nt},\; \Theta=\lambda \sum_{k} L_{k}e^{-i\lambda kt}
\]
we get a (classical) Virasoro algebra
\[
[L_{m},L_{n}]= -i \{ (m-n)L_{m+n}+\eta_0 m^3 \delta_{m+n} \}
\]
Hence it's shown that the (quantum) central charge is $\mathcal{C}=
12\eta_0$. With one qualification, this is the same result as obtained
in \cite{nns}. The qualification to be added is that the authors of
\cite{nns} refer to a non scalar dilaton, here we improve upon the
scalar dilaton theory studied in \cite{cm2}.\\
Turning to the issue of the statistical entropy, it is given by
Cardy's formula, provided that \(a^2\gg 24\),
\[
S_{stat}=2\pi\sqrt{\mathcal{C}L_{0}/6}
\]
where \(L_{0}=M/\lambda \). Substituting for the value of the
central charge 
\[
S_{stat}=2 \pi \eta _{0}a
\]
which agrees with the thermodynamical expression (\ref{tdentr}). This
demonstrates also that, at least for 2D systems, discarding boundary
terms leads to wrong conclusions. This is probably related to the
topological nature of 2D gravity. As a remark, for
higher dimensional AdS black holes\footnote{These are described
in \cite{hp83,abhp96,mann97,van97,blp97,bir99}.}, the horizon central
charge (calculated a l\`a Carlip\cite{car99}) matches the
Bekenstein-Hawking entropy provided the mass of the black hole, but
not the mass as measured from infinity, be directly related to the
Virasoro's generator $L_0$. On the other hand, the
AdS\(_{d+1}\)/CFT\(_{d}\) correspondence shows that the
entropy scales with the area of the event horizon, so the CFT at
infinity with a black hole in the bulk must count correctly the number
of degrees of freedom of the horizon (see \cite{ks91} for a discussion of
density of states in CFT), in agreement with Witten's argument 
\cite{wit98}. \\
Our result removes one of the problems in the AdS\(_{2}\)/CFT\(_{1}\)
scenario but also introduces new questions. Here we show that
asymptotic symmetries correctly account for the entropy of the two
dimensional black holes; on the other hand we recently showed that
regularity at the horizon also do the job \cite{ccv}. As for $AdS_3$,
this seems to suggest that there is some correspondence between states
at infinity and on the horizon in 2D systems. But unlike from $AdS_3$,
the central charge here is equal to the one calculated using near
horizon symmetries, and in fact the $AdS_2$ black hole simply is a
foliation of $AdS_2$ by a different choice of time coordinate.

\end{document}